\documentstyle[stwol,epsfig,rotating]{article}

\def\Journal#1#2#3#4{{#1}{\bf #2} (#4) #3}
\def\ACPB{{\em Acta. Phys. Pol.   }{\bf B}}
\def\CPC{\em Comp. Phys. Comm.    } 
\def\JETP{\em Sov. Phys. JETP     }
\def\MPLA{{\em Mod. Phys. Lett.   }{\bf A}}
\def\NCA{\em Nuovo Cimento        }

\def\NPB{{\em Nucl. Phys.         }{\bf B}}
\def\PLB{{\em Phys. Lett.         }{\bf B}}
\def\PRL{\em Phys. Rev. Lett.     }
\def\PR0{\em Phys. Rev.           }
\def\PRD{{\em Phys. Rev.          }{\bf D}}
\def\PRP{\em Phys. Rep.           }

\def\SJNP{\em Sov. J. Nucl. Phys. }
\def\ZPC{{\em Z. Phys.            }{\bf C}}
\def\ra{\rightarrow}
\def\pp{\mbox{$\bar{p}p$~}}
\def\ee{\mbox{$e^+e^-$~}}
\def\eea{\ee anni\-hi\-la\-tion}
\def\gp{\gamma p}
\def\ptm{p_{t,\mbox{\scriptsize max}}}
\def\beq{\begin{equation}}
\def\eeq{\end{equation}}
\def\avE{\langle E \rangle}
\def\avn{\langle n \rangle}
\def\Pom{I\!\!P}
\def\xP{x_{\Pom}}                                                           
\def\aP{\alpha_{\Pom}}                                        
\def\aPP{\alpha_{\Pom}^{\prime}}
\def\xPs{x_{_{\Pom}}}

\def\F2D{F_2^{D(3)}}
\begin{document}
%
%
\begin{titlepage} 
\begin{large}
\begin{flushright} X--LPNHE/96--10 \\ November  1996 \\ \end{flushright}
\end{large} 
\vspace*{40mm}         
\begin{center}
{\LARGE \bf               Soft Interactions \\ 
\vskip  7pt            and Diffraction Phenomena} \\    
\vskip 30pt            {\Large \bf S.V. Levonian} \\
\begin{large}
\vskip 12pt  {\em LPNHE, Ecole Polytechnique, F-91128 Palaiseau, France \\
                     and LPI RAS, 117924 Moscow, Russia}   \\
\vskip 5pt         
                    {\tt e-mail: levonian@polhp2.in2p3.fr  \\
                    http://www-h1.desy.de/$\sim$levonian}  \\
\vspace*{20mm}       
                                Plenary talk \\
\vskip -2pt     given at the 28$^{th}$ International Conference 
                          on High Energy Physics, \\              
                         Warsaw, 25-31 July, 1996
\end{large}
\end{center}
\vskip 80pt
\begin{abstract}
\noindent
    Until the mystery of confinement is understood from the first
    principles, so called {\em soft physics} remains an important 
    area of research, providing valuable information on  
    underlying dynamics of strong interactions at long distances.
    In this review an attempt is made to summarize recent
    experimental results on  multiparticle production
    in \eea~ and on diffraction at HERA.
\end{abstract}
\end{titlepage}
\newpage \thispagestyle{empty} \mbox{ } \newpage
\addtocounter{page}{-1}
%
%
\title{SOFT INTERACTIONS AND DIFFRACTION PHENOMENA}

\author{ S.V. LEVONIAN }

\address{LPNHE, Ecole Polytechnique, F-91128 Palaiseau, France \\
               and LPI RAS, 117924 Moscow, Russia}

\twocolumn[\maketitle\abstracts{
    Until the mystery of confinement is understood from the first 
    principles, so called {\em soft physics} remains an important 
    area of research, providing valuable information on  
    underlying dynamics of strong interactions at long distances.
    In this short review an attempt is made to summarize recent
    experimental results on  multiparticle production
    in \eea~ and on diffraction at HERA.}]

\section{Introduction}
Contrary to a widely spread prejudice that soft physics at high 
energies is just a boring obstacle in the way of fast progress,
I want to argue that it remains of a fundamental importance, providing us
with new and valuable information about the underlying dynamics of strong
interactions.
 
There are at least two basic reasons to study  so called 
{\em soft processes} in particle physics.
First of all, they represent the most fundamental challenge 
in QCD~\cite{Gross}. This is in contrast to the perturbative sector (pQCD),
where the major theoretical problems are rather of a technical nature 
(to carry out difficult, but well defined calculations).
Secondly, even if one is interested in  precision pQCD tests, 
it is often difficult -- if not impossible --
to disentangle various non-perturbative effects.
It is not only the unavoidable fragmentation phase, but also the presence
of soft underlying events~\cite{in_SUE}, originating from multiple
parton-parton scattering, which can mask the properties of
a hard subprocess.
Therefore, until the nature of confinement is understood theoretically 
from first principles, experimental data which can shed  additional light
on the problem remain very important.

There exists a large variety of physics phenomena where
non-perturbative QCD effects are essential. 
In this review three classes of such phenomena are discussed, 
for which new experimental results have been reported at the conference.
These are: 
     multiparticle production in \eea,
     diffractive physics at HERA and 
     the transition from soft to hard scattering in $ep$ collisions.
Theoretical aspects of diffraction in QCD will be discussed separately,
by George Sterman in his plenary talk~\cite{Sterman}.

\section{Multiparticle production}   \label{sec:two}
In this domain  information about underlying production mechanisms
is extracted by studying the correlations among produced particles.

A major difficulty in hadron-hadron collisions arises from the fact, 
that several non-perturbative effects get mixed here.    
Therefore,  hard collisions are important to study the properties
of hadronization, the only non-perturbative phase in the process. 
The details of the fragmentation  at low scale, $Q_0$ 
can shed a light on the preconfinement stage 
(i.e. the dynamics between $Q_0$ and $\Lambda_{QCD}$) followed by the 
recombination of quarks and antiquarks into real hadrons.

$e^+e^-$ collisions with their ``clean QCD environment''
provide the best laboratory for such kind of measurements.
A typical strategy here is to calculate analytically the partonic cascade
down to a scale $Q_0$ and then apply a phenomenological 
hypothesis, like e.g. local parton-hadron duality~\cite{LPHD},
to end up with observable hadron distributions which can be 
confronted with experimental data.

When the basic properties of parton-to-hadron fragmentation are established,
one can go back to a difficult case of hadron-hadron collisions,
to learn more about non-perturbative dynamics there.

\subsection{Quark and Gluon Jet Properties} \label{subsec:qgj}
One of the specific QCD predictions is a difference between the properties
of quark and gluon initiated jets. This effect is due to the different 
relative probabilities for a gluon or a quark to radiate an additional gluon,
given by the Casimir factors $C_A=3$ and $C_F=4/3$ respectively.
Experiments~\cite{LEP_old} confirm, at least at the qualitative level, that
gluon jets are ``richer'': 
they  have a softer and significantly broader fragmentation function and they
are characterized by larger average multiplicity, as compared to quark jets. 
To study this effect the experiments at LEP use 
high statistics (few $10^6$ hadronic $Z_0$ decays)
and their excellent ability to tag jet flavour (with a typical purity of 
$\sim 80\div90\%$) in the clean QCD environment of \eea.

Especially interesting is to study the ratio
        $$  r \equiv \avn_{gluon} / \avn_{quark}  $$
for which analytical QCD calculations exist.
From the theoretical side, the progress over the last two decades
in computing the ratio $r$ resulted in decreasing its value from
$r=C_A/C_F=9/4$ in leading order QCD~\cite{f_bg} 
to $r = 1.84 \pm 0.02$ in higher order QCD,
taking into account recoil effect in the soft gluon emission~\cite{f_dh}.
Experiments however found lower values, concentrated around $r \approx 1.2$.

A word of caution is in order here.
Whereas the theoretical calculations refer to the parton multiplicities
originating from $q\bar{q}$ or $gg$ states, experiments measure
charged particles, applying in addition specific jet algorithms to select 
three jet events $e^+e^- \ra q\bar{q}g$ as a source of gluon jets.
Hence the experimental results depend upon the jet algorithm
chosen~\cite{LEP_old}.

It was demonstrated~\cite{f_dh,gary} 
with the help of Monte Carlo simulations, that hadronization, 
when taken into account, brings the theoretical value
closer to those measured experimentally.
In order to eliminate a mismatch between the
theoretical and experimental definitions of a gluon jet, 
a new method was proposed~\cite{gary}, based upon  rare topologies
of $e^+e^-\!\ra\!q\bar{q}g$ events, 
in which both quark jets appear in the same hemisphere
of an event and hence are balanced by the gluon.
In this case the multiplicity $\avn_{gluon}$
can be determined without applying a jet algorithm, 
just by counting the particles in the `gluon hemisphere'.

New results presented at this conference~\cite{LEP_new},
are based on yet higher statistics as well as improved particle 
identification and flavor tagging technique.
DELPHI has measured $r=1.16 \pm 0.01$ at $\avE=24$ GeV and 
$r=1.28 \pm 0.02$ at $\avE=30$ GeV, using a natural mixture of all quark
flavours in their sample ($q=udscb$).

OPAL, motivated by the fact that massless quarks are used in theoretical
calculations, made a step further by separating light and heavy quarks 
and has shown, that
   $$ \langle n_{ch}\rangle(uds) < 
      \langle n_{ch}\rangle(b)   <
      \langle n_{ch}\rangle(g),  $$
leading to different values of
$r=1.09 \pm 0.03$ for ($q=b$) and $r=1.39 \pm 0.05$ (for $q=uds$),
at $\avE=24$ GeV.

In yet another OPAL analysis~\cite{OPAL2} a full use of all available 
statistics was necessary in order to exploit the new method of 
Gary~\cite{gary}, described above.
Only 221 out of $3.04\cdot 10^6$ events were selected for the
final `gluon-sample'.
As a result, a value of
             $$ r=1.575 \pm 0.046_{stat} \pm 0.069_{syst} $$
was obtained, for the average energy of $\avE=39$ GeV. 
\begin{figure}[t]
 \center
  \begin{picture}(78,105)(1,0)
  \epsfig{figure=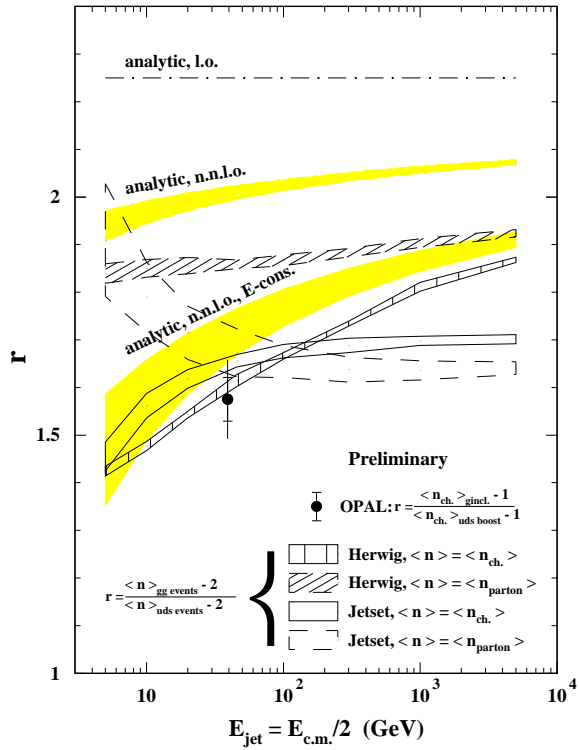,width=78mm,%
   bbllx=90pt,bblly=240pt,bburx=450pt,bbury=655pt}
  \end{picture}
  \caption{QCD analytic and Monte Carlo predictions for the ratio 
           of the mean particle multiplicity between gluon and quark
           jets as a function of jet energy, in comparison to the OPAL
           measurement for 39 GeV jets~\protect\cite{OPAL2}.} 
 \label{fig:OPAL}
\end{figure}
As one can see from Fig.~\ref{fig:OPAL} this value is already in a very
good agreement with both analytic calculations and  Monte Carlo 
predictions.\footnote
  {There was also new CLEO result~\cite{CLEO}, presented at the 
   conference: 
   $ r = 1.04 \pm 0.02(\mbox{stat}) \pm 0.04(\mbox{syst}), $
   obtained for jet energies less than 7 GeV.
   For the correct interpretation however one has to treat 
   threshold effects properly.
   It seems there is simply no phase space available 
   to develop a fat partonic cascade at this low energy.}

\subsection{Multiplicity Distributions} \label{subsec:md}
Multiplicity distributions (MD), i.e. a set of probabilities, $P_n$ 
to produce $n$-particle final state, are natural and most widely used
characteristics of multiparticle production.
The state of the art in this field can be found 
in recent reviews~\cite{MD1,MD2}.
In this section two questions are briefly discussed:
\begin{itemize}
  \item  do we understand the basic properties of MD at least
         in hard scattering, taking as an example the process
         $e^+e^-\!\ra\! Z_0\!\ra\! hadrons$?
  \item  is there an analogy between MD in hard (\ee) and
         predominantly soft ($hh$) collisions?
\end{itemize}

Apart from well known average multiplicity, $\avn$, and dispersion, $D$,
other observables are used in  modern analyses of MD~\cite{MD1}:
factorial ($F_q$) and cumulant ($K_q$) moments of rank $q$, and,
since recently, their ratio, $H_q = K_q/F_q$. 
 These quantities are helpful in the mathematical description of branching
 processes, and they are widely used in various scientific fields
 to describe the properties of cascade phenomena~\cite{Bialas}.
It has been shown~\cite{MD_Hq} that the ratio $H_q$ is not only sensitive
to the shape of MD, but also can be calculated in higher-order pQCD.

Phenomenologically, the best description of the MD in \eea~
is provided by the negative binomial (NBD) or the log-normal (LND)
distributions~\cite{MD1}. Even they however fail to reproduce 
all the details observed in the data. 
In particular, experimental data show two interesting
features~\cite{MD_sh,SLD}: \\
\indent  1) a pronounced shoulder in MD in the range of intermediate 
            multiplicities and \\
\indent  2) a quasi-oscillatory behavior of the ratio $H_q$ as a function 
            of rank $q$.

\begin{figure}[t]
 \center
  \begin{picture}(78,101)(1,0)
  \epsfig{figure=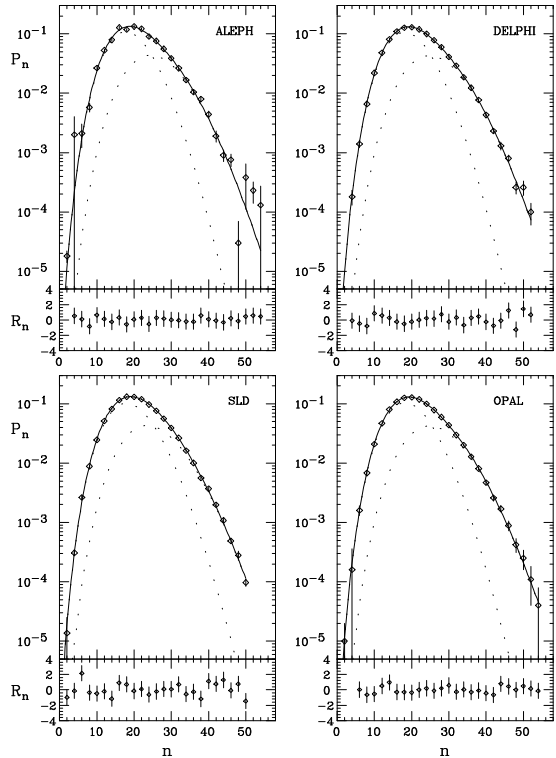,width=78mm,%
   bbllx=70pt,bblly=220pt,bburx=480pt,bbury=800pt}
  \end{picture}
  \caption{Charged particle multiplicity distributions, $P_n$,
           measured at the $Z_0$ peak, as compared to 
           the superposition of two NBD contributions (solid curves),
           which are also shown separately (dotted curves).
           The residuals, $R_n$, are shown underneath each of the 
           distributions.}
 \label{fig:MD1}
\end{figure}

\begin{figure}[t]
 \center
  \begin{picture}(78,90)(1,1)
  \epsfig{figure=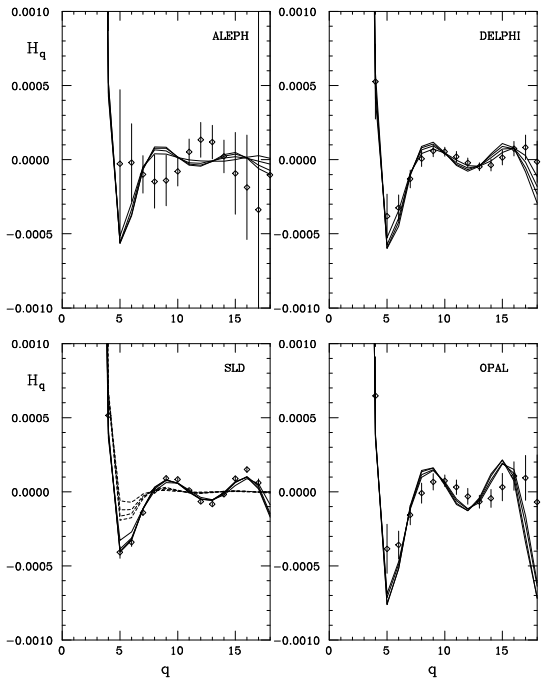,width=78mm,%
   bbllx=30pt,bblly=220pt,bburx=530pt,bbury=820pt}
  \end{picture}
  \caption{The ratio $H_q$ as a function of the rank, $q$.
           The data are compared to the superposition of two NBD,
           weighted according to the relative proportion of 2- and
           multi-jet events, as measured in each of the shown experiments.
           The difference between the dashed and the solid lines at
           SLD plot shows the effect of the truncation of MD.}
 \label{fig:MD2}
\end{figure}

In the paper~\cite{pa02_6} presented to this conference, authors attempted
to find a common origin of both aforementioned effects.
They have shown (see Fig.~\ref{fig:MD1} and Fig.~\ref{fig:MD2})
that good quantitative description of the data can be obtained
by a superposition of two NBD's with the relative proportion given
by the ratio of 2-jet to multi-jet events as defined in the same experiments.
It was also demonstrated that the truncation effect
($n_{ch} < n_{max}$) in experimentally measured MD cannot be neglected.  

Thus, the shape of MD in \eea~ can be understood in terms of the
multi-jet event topology, predicted by pQCD.

Let us now turn to the ``dirty''world of  $hh$-collisions.
Here the bulk of the cross section is due to soft, low $p_t$ processes,
and hence pQCD approach is not fully justified.
Surprisingly, we find nevertheless, that both discussed features of MD
are well familiar in $hh$ experiments as well~\cite{MD_hh1,MD_hh2}.

As we have seen, an essential ingredient for a successful description
of the data is a {\em two-component} structure of MD.
In hard collisions (\ee) it could be identified with two- and
multi-jet events.
What is the origin of these components in predominantly soft 
$hh$ processes?
Looking  further back in history, we do find a good candidate.
It was shown that the Dual Parton Model (DPM)~\cite{DPM}, 
based on Reggeon field theory~\cite{RFT}, is able to reproduce 
the shape of KNO-distributions and to describe
two-particle correlations in high energy $pp$- and \pp-collisions.
In the framwork of DPM, multi-pomeron exchanges, which are necessary 
to unitarize the hadron-hadron scattering amplitude, naturally lead  
to the different event topologies, consisting of 1-, 2-, ... 
Pomeron strings.

It would be very interesting to see whether the oscillations of $H_q$
are also reproduced by DPM.

\section{Diffraction and Physics of Pomeron} \label{sec:three} 
Another large area where soft physics plays the dominant r\^{o}le
is diffraction.
The basic properties of diffractive processes 
-- cross sections exponentially falling 
with four momentum transfer squared, $t$,
predominantly low mass final states, $M^2\!\!<\!\!<\!\!s$,
large rapidity intervals between the diffractively produced hadronic
systems, etc. -- have been established
already in the 60's and 70's in hadron-hadron scattering experiments.
These properties are typical for soft, peripheral interactions, and
they found a best and most natural  explanation in 
a framework of Regge approach~\cite{Regge,DD_RR}. 
A number of comprehensive review articles~\cite{DD_R1}$^-$\cite{DD_R3}
nicely summarise the experimental status in this
field and the interpretation of the data in terms of Regge theory.
 
Since the advent of the parton model and then QCD
a main emphasis was shifted to {\em hard processes}, leaving for
diffraction only a little attention.
A spike of interest was caused by the observation of jets in 
\pp diffractive final states~\cite{DD_UA8}, following the pioneering
proposal by Ingelman and Schlein~\cite{DD_IS}, who invented a concept
of the Pomeron structure function.
A real ``Renaissance'' however was triggered by early HERA data~\cite{F2}
on the proton structure function, $F_2$ at low $x$,
and the observation of large rapidity gap events in deep inelastic $ep$
scattering~\cite{HLRG}.
Only then it was clearly realized, that diffraction and 
low $x$ physics at HERA are closely related subjects with high potential
to provide a completely new insight into non-perturbative QCD effects, 
as well as to bridge the soft (Regge) and hard (pQCD) domains.

\subsection{Total Cross Sections and Pomeron} \label{subsec:tot}
\begin{figure}[t]
 \center
  \begin{picture}(80,105)(15,2)
   \begin{sideways} 
    \put(83,-8){$\sigma_{\bar{p}p,pp} /~$mb} 
    \put(52,-8){$\sigma_{\gamma p} /~\mu$b}
    \put(23,-8){$\sigma_{\gamma\gamma} /~$nb} 
    \end{sideways}
    \put(40,3){$E_{_{CM}}/$GeV}
    \epsfig{figure=F_04.eps,width=90mm,angle=90}
    \put(-41,33){\epsfig{file=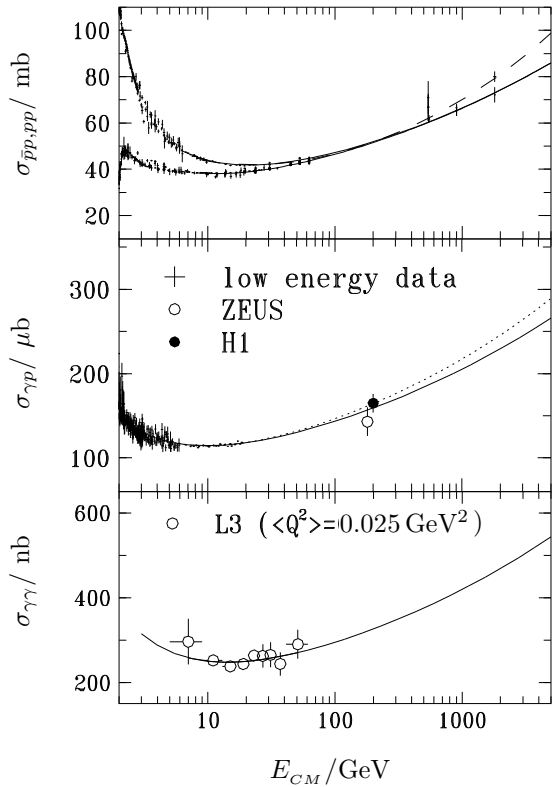}}
    \put(-40.7,35.7){$0.025\,$GeV$^2$} \put(-23,36){)}
   \end{picture}
  \caption{Energy dependence of the total $pp, \bar{p}p, \gamma p$ and 
         $\gamma\gamma$ cross sections. The curves represent the DL
         parametrization~(\protect\ref{eq:Sigma})
         with $\lambda=0.0808$ (solid), $\lambda=0.112$ (dashed) and
         $\lambda=0.088$ (dotted).}
 \label{fig:stot}
\end{figure}

In Regge theory the asymptotic behavior of the energy dependences
of total, elastic and diffractive dissociation cross sections
is defined by the properties of the leading vacuum (Pomeranchuk)
trajectory:
\beq   \aP(t) = 1 + \lambda + \aPP t   \label{eq:sP}      \eeq  
with the intercept $\aP(0) \equiv \Delta = 1 + \lambda$.
As has been demonstrated by Donnachie and Landshoff~\cite{d_DL}, 
the energy dependence of {\em any} total hadronic cross section 
$\sigma_{ab}$ can be described in a very economical form   
\beq 
   \sigma_{ab} = X_{ab} \cdot s^{\lambda} + Y_{ab} \cdot s^{-\eta}
\label{eq:Sigma}
\eeq
with universal powers
\beq 
 \lambda \approx 0.0808, ~~~  \eta \approx 0.4525  \label{DLfit} 
\eeq
being independent of the type of colliding particles  $a$ and $b$ 
(Fig.~\ref{fig:stot}).
This simple Regge inspired parametrization, when applied to
photoproduction, successfully predicted $\sigma_{\gp}$ at HERA energies.
At this conference new data have been presented by the L3 Collaboration
\cite{d_L3} on total gamma-gamma cross sections at small
virtualities $\langle Q^2 \rangle = 0.025$ GeV$^2$. 
The data have been extracted from \ee collisions at $\sqrt{s}=130, 136$
and 140 GeV, recently collected at LEP on its way to LEP-200.  
As one can see in Fig.~\ref{fig:stot} these data are
also consistent with the parametrization~(\ref{eq:Sigma}).
The dashed and the dotted curves on Fig.~\ref{fig:stot}
demonstrate the energy dependence~(\ref{eq:Sigma}) with $\lambda=0.112$
and $\lambda=0.088$ respectively, obtained separately from the 
best description of the Tevatron data~\cite{d_TevP} and of photoproduction
data including HERA measurements~\cite{HERA_stot} into the fit.
It should be stressed however, that the original result~(\ref{DLfit})
remains a good {\em global} fit of all total hadronic cross sections,
leading to a very attractive concept of a {\em universal Pomeron}. 

In all so far discussed reactions  soft, long distance physics represents
the dominant contribution.  
Therefore, the value of $\Delta \approx 1.08$ is often referred to
as the intercept of the `soft' Pomeron.
In case of deep inelastic scattering (DIS) where it is natural to expect
hard processes to dominate, a much steeper rise was observed  
of $\sigma_{tot}(\gamma^*p)$ with energy $W$,
(or, equivalently, a rise of $F_2$ with decreasing Bjorken $x$)
which can be
approximated by a power low $(W^2)^{\lambda}$ with $\lambda \approx 
0.2 \div 0.4$ depending on the virtuality $Q^2$ of the photon~\cite{GAb}. 
At the beginning it was tempting to explain this fast rise as a 
manifestation of perturbative BFKL Pomeron~\cite{BFKL}, which in leading 
logarithmic approximation (LLA) corresponds to 
\beq 
\lambda_{_{BFKL}} = \frac{4N_c \ln 2}{\pi} \alpha_s \approx 0.4 \div 0.5
\label{eq:BFKL}
\eeq
This value of $\lambda_{_{BFKL}}$, as was pointed out by the authors
themselves~\cite{NLO_BFKL}, should not be seen uncritically, 
since next-to-leading order corrections
may significantly change the result of  LLA calculation.
Therefore, more realistic and cautious expectation for the intercept 
of the `hard' Pomeron (sometimes also referred to as the 
{\em QCD Pomeron}) is  $\lambda \geq 0.2 \div 0.3$.

Reaching this stage, an innocent reader may immediately conclude:
$$ 0 < \lambda_{soft} < \lambda_{hard} < \lambda_{_{BFKL}} \leq 0.5 $$
and wonder, how many pomerons are there after all?
Moreover, whatever the exact value of $\lambda$ is, 
eq.~(\ref{eq:Sigma}) will sooner or later violate the Froissart bound.

The variety of the opinions can be summarized in the following way.
First of all, everybody agrees, that unitarity corrections to 
one-pomeron exchange are necessary, 
although different technical schemes were proposed
for the unitarization procedure:
multi-pomeron exchanges~\cite{sd_CKMT}, eikonalization of the scattering
amplitude~\cite{sd_GLM}, or the combination of both~\cite{sd_MS}.
Thus, the intercept of the `bare' Pomeron may be significantly different
from the `effective' intercept measured in the experiments, 
depending on how large unitarity corrections 
to a single-pomeron exchange amplitude are. 
At this point the opinions (models) start to diverge: \\
\indent
1) At presently available energies unitarity corrections are negligible,
   and the universal `bare' Pomeron has an intercept of 
   $\Delta \approx 1.08$.
   These models are in fact ruled out by  $F_2$ measurements at HERA. \\
\indent
2) The universal (i.e. the same in hadron-hadron and in DIS reactions) 
   `bare' Pomeron has an intercept of $\Delta \approx 1.25$.
   The effect of screening (unitarity) corrections depends 
   on both the energy and   a typical hard scale involved. 
   The value of $\Delta_{eff}=1.08$ observed in hadronic reactions
   is thus a result of large corrections, whereas in DIS
   the corrections become small with increasing $Q^2$, and hence
   the `bare' Pomeron is seen. \\
\indent
3) `Bare' Pomeron $\equiv$ BFKL Pomeron, but screening corrections are
   so large that the effective power $\lambda$ becomes smaller
   even in the DIS case. \\
\indent
4) There is no Pomeron whatsoever in DIS.
   The rise of $F_2$ at small $x$ can be equivalently well described
   by the conventional DGLAP~\cite{DGLAP} evolution equations,
   while the rapidity gap events in DIS may be due to various  
   mechanisms~\cite{nonP}, which do not need the concept of Pomeron.
      
Further, more illuminating discussion about the `soft' and `hard'
Pomeron can be found elsewhere~\cite{p95_jb}.
From our point of view, there is only one Pomeron
(and in that sense it is universal),
which however receives contributions from both non-perturbative and
perturbative exchanges. Depending on the experimental conditions,
either soft, or hard component will dominate, thus leading to the 
different values of $\lambda$ observed. Screening corrections may
additionally reduce the value of the `effective' intercept. 
 
As already was mentioned (for details see Ref.~\cite{Sterman,GAb}),
$F_2$ measurement alone is too inclusive to conclude unambiguousely
on BFKL dynamics.
Therefore, dedicated measurements of diffractive processes are 
necessary to decide whether or not `hard' Pomeron is seen at HERA.
We now turn directly to this topic.

\subsection{Large Rapidity Gap Events and Diffraction} \label{subsec:LRG}
\begin{figure}[t]
 \center
  \begin{picture}(80,46)(-5,0)
   \epsfig{figure=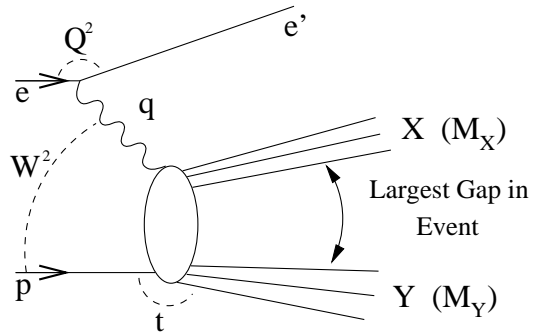,width=70mm,angle=270}
  \end{picture}
  \caption{Generic diagram for a rapidity gap event in $ep$ scattering.}
 \label{fig:RG}
\end{figure}

Experimentally, diffractive processes at HERA  are studied by looking for
so called {\em Large Rapidity Gap} events (Fig.~\ref{fig:RG}),
which are naturally expected, and are not exponentially suppressed,
in the case of colourless exchange between the
proton and the virtual photon.
Defining as usual $Q^2=-q^2$ and $W^2=(q+p)^2$, where $q$ and $p$ denote
the 4-vectors of the virtual photon and the proton respectively, 
further relevant kinematic variables can be introduced:
$$
   t=(p-Y)^2,~~
   \xP=\frac{Q^2+M_{_X}^2}{Q^2+W^2},~~
   \beta=\frac{Q^2}{Q^2+M_{_X}^2}.
$$
Note, that despite the suggestive names (e.g. $\xP$) the described
experimental definition is completely general.  
A cross section measured at different values of the aforementioned
kinematical variables is free of any assumptions on the nature
of contributing processes.
It is only when interpreting the results in terms of diffractive
dissociation that one has to select a kinematical range where the 
Pomeron exchange dominates (small $\xP \leq 0.05$)
and to take into account possible remaining non-diffractive contributions.

The H1 and ZEUS experiments use the capabilities of their forward detectors
to either identify the leading baryonic system $Y$ as a proton
or a neutron, or to restrict its mass, $M_{_Y} < 1.6 \div 2$ GeV, 
if the system $Y$ is not directly observed.
This guarantees the gap between the hadronic systems $X$ and $Y$
to be larger than 3 to 4 units in rapidity.
Furthermore, the specifically asymmetric kinematics of HERA, 
due to very different proton and lepton beam energies,
ensures a uniquely high detector acceptance for the diffractively produced
hadronic system $X$. This enables precision measurements of the
diffractive dissociation of virtual photon at HERA.
Finally, high resolving power ($Q^2$) of the virtual photon
allows for the first time in high energy physics the 
partonic sub-structure of the exchanged object in diffraction to be probed.

\subsection{Soft Diffractive Processes} \label{subsec:sd}
A special interest to soft diffractive processes is 
related to the expectation to see an effect of unitarity corrections
to the diffractive cross sections at presently available energies.
Indeed, since a Pumplin limit~\cite{d_Pump}, 
 $\sigma_{el}/\sigma_{tot} \leq \frac{1}{2}$, is violated 
significantly earlier than the Froissart bound~\cite{d_Frois}, 
$\sigma_{tot} < c\ln^2s$,
one expects unitarity corrections to a single pomeron exchange to
become more important in diffractive processes compared to the total
cross section.
First experimental evidence of this has been reported for \pp collisions 
at the Tevatron~\cite{sd_CDF}.

It is well known in Regge phenomenology that the information about the
Pomeron intercept can be extracted not only from the energy dependences
of the total and diffractive cross sections, but also from the diffractive
mass distributions. 
Moreover, in the context of eikonal models~\cite{sd_GLM}
the differential cross section 
$d\sigma^D/dM_{_X}^2 \propto (M^2)^{-(\Delta+2\aPP t)}$
is not affected by screening corrections. 
Hence a 'bare' Pomeron intercept is expected to be seen.
Thus comparing the values of $\Delta$ extracted from the diffractive mass
distributions with those obtained from the energy dependence of the
cross sections may provide important information about screening
corrections in soft diffractive processes.
  
At this conference two HERA experiments have presented first preliminary 
data on inclusive diffractive photoproduction at comparable energies of
$\langle W_{\gp}\rangle=187$ GeV (H1~\cite{sd_H1}) and
$\langle W_{\gp}\rangle=200$ GeV (ZEUS~\cite{sd_ZEUS}).
%
\begin{figure}
 \center
  \begin{picture}(80,78)(0,1)
  \epsfig{figure=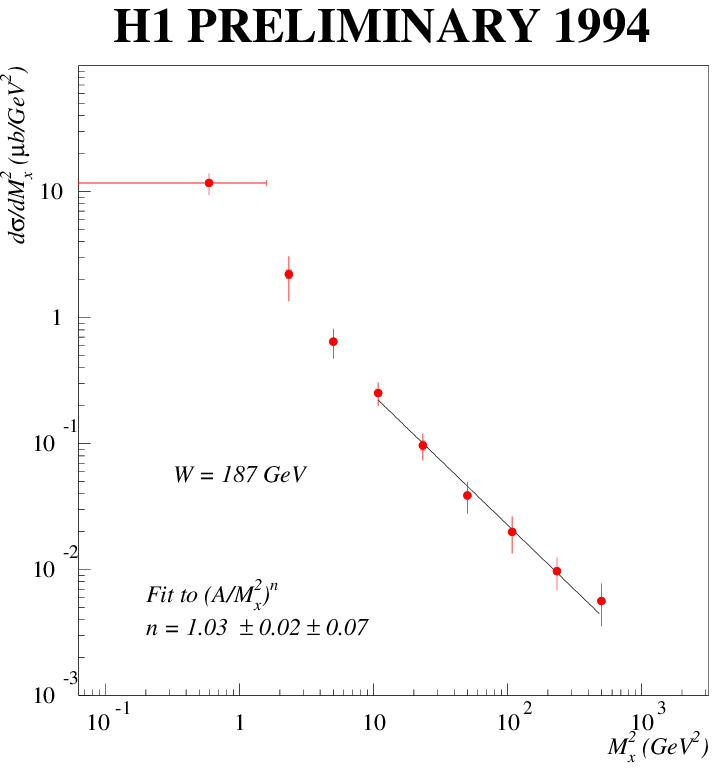,width=80mm}
  \end{picture}
  \caption{The differential cross section, $d\sigma/dM_{_X}^2$
         for the process $\gamma p \rightarrow XY$ in the kinematic
         region $M_{_Y} < 1.6$ GeV and $|t| < 1$ GeV$^2$.
         A fit is shown for $10.8 < M_{_X}^2 < 500$ GeV$^2$.}
 \label{fig:gpMx}
\end{figure}
Triple Regge analyses of the diffractive mass spectra in the
reaction $\gp \ra M_{_X}p$ (see Fig.~\ref{fig:gpMx}),
after correcting for the unmeasured $t$-dependences, yield
the intercept values, entirely consistent with the `soft' Pomeron:
$$ \begin{tabular}{r l}
     H1: &~ $\Delta=1.11\pm 0.02_{stat}\pm 0.07_{syst}$, \\
   ZEUS: &~ $\Delta=1.14\pm 0.04_{stat}\pm 0.08_{syst}$. \\ 
\end{tabular} 
$$
Yet the data and their precision are insufficient for any firm conclusion
about the importance of screening corrections at HERA energies.

\subsection{Hard Diffraction at HERA} \label{subsec:hd} 
\begin{figure}
 \center
  \begin{picture}(78,133)(1,5)
   \epsfig{figure=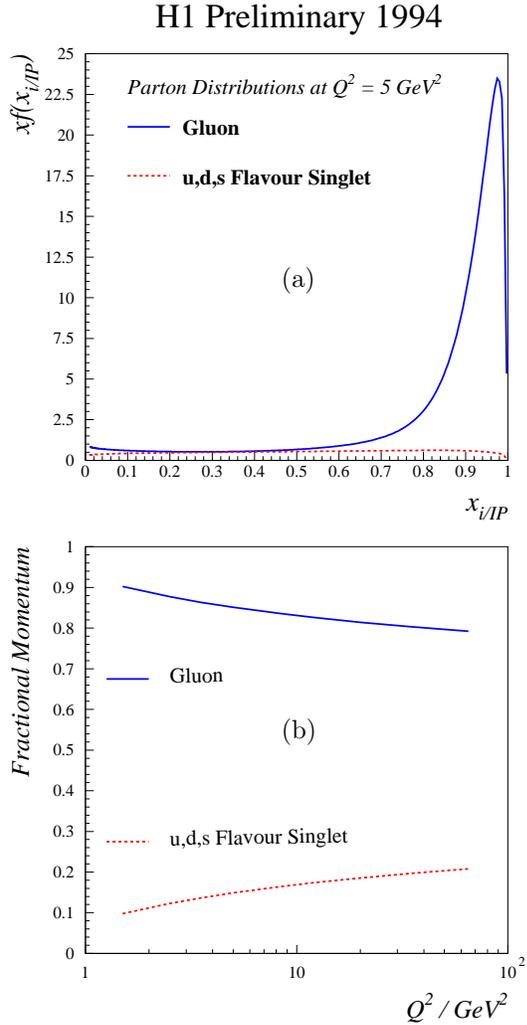,width=78mm,angle=270}
   \put(-40,104){(a)}  \put(-40,44){(b)}
  \end{picture}
  \caption{(a) Partonic content of the Pomeron at the starting scale, 
               $Q^2 = 5$ GeV$^2$ resulted from  
               QCD fit~\protect\cite{H1_F2D}
               of the diffractive structure function $\F2D$;
           (b) fraction of the total momentum in diffractive exchange
               carried by quarks and by gluons, as a function of $Q^2$.}
 \label{fig:H1QCD}
\end{figure}

First direct information about the partonic structure of the Pomeron
can be extracted from the inclusive measurements in diffractive DIS:
\beq
  \frac{\mbox{d}^3\sigma^D}{\mbox{d}\beta dQ^2 \mbox{d}\xPs} =
  \frac{4\pi\alpha^2}{\beta Q^4}(1-y+\frac{y^2}{2}) \F2D(\beta,Q^2,\xPs),
\label{eq:F2D3}
\eeq
where $y=Q^2/(sx)$ is the inelasticity parameter.
New data on the differential cross section~(\ref{eq:F2D3}) have been
presented by ZEUS~\cite{Z_F2D} and H1~\cite{H1_F2D}.
High statistics, based on 2 pb$^{-1}$ of 1994 data, enables H1 to perform
QCD analysis of the diffractive structure function $\F2D$
which resulted in the parton distributions, shown in Fig.~\ref{fig:H1QCD}. 
As one can see more than $80\%$ of the momentum
in the diffractive exchange is carried by gluons.
Important technical details of this interesting analysis are described 
in Ref.~\cite{H1_F2D}. 

\begin{figure}
 \center
  \begin{picture}(75,25)(-2,0)
   \epsfig{figure=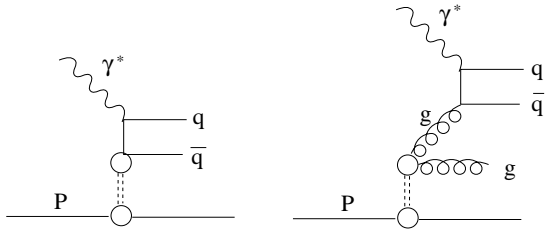,width=72mm,angle=270}
  \end{picture}
  \caption{Feynman diagrams for quark and gluon initiated processes
           in diffractive DIS.}
 \label{fig:FD}
\end{figure}

Such a `gluon dominated Pomeron' leads to definite predictions about 
the properties of the hadronic final states in DIS diffraction,   
which can be confronted with experimental data.
For example, as illustrated by Fig.~\ref{fig:FD}, 
more energy flow, bigger $\langle p_t \rangle$ and lower thrust values
in the $\gamma^*\Pom$ centre of mass system are expected
in case of a gluon dominated exchanged object, as 
compared to a quark dominated one. 
A significant difference is also expected in the  production rates
of charm and of high $E_t$ jets.

\begin{figure}[t]
 \center
  \begin{picture}(80,95)(0,0)
   \epsfig{figure=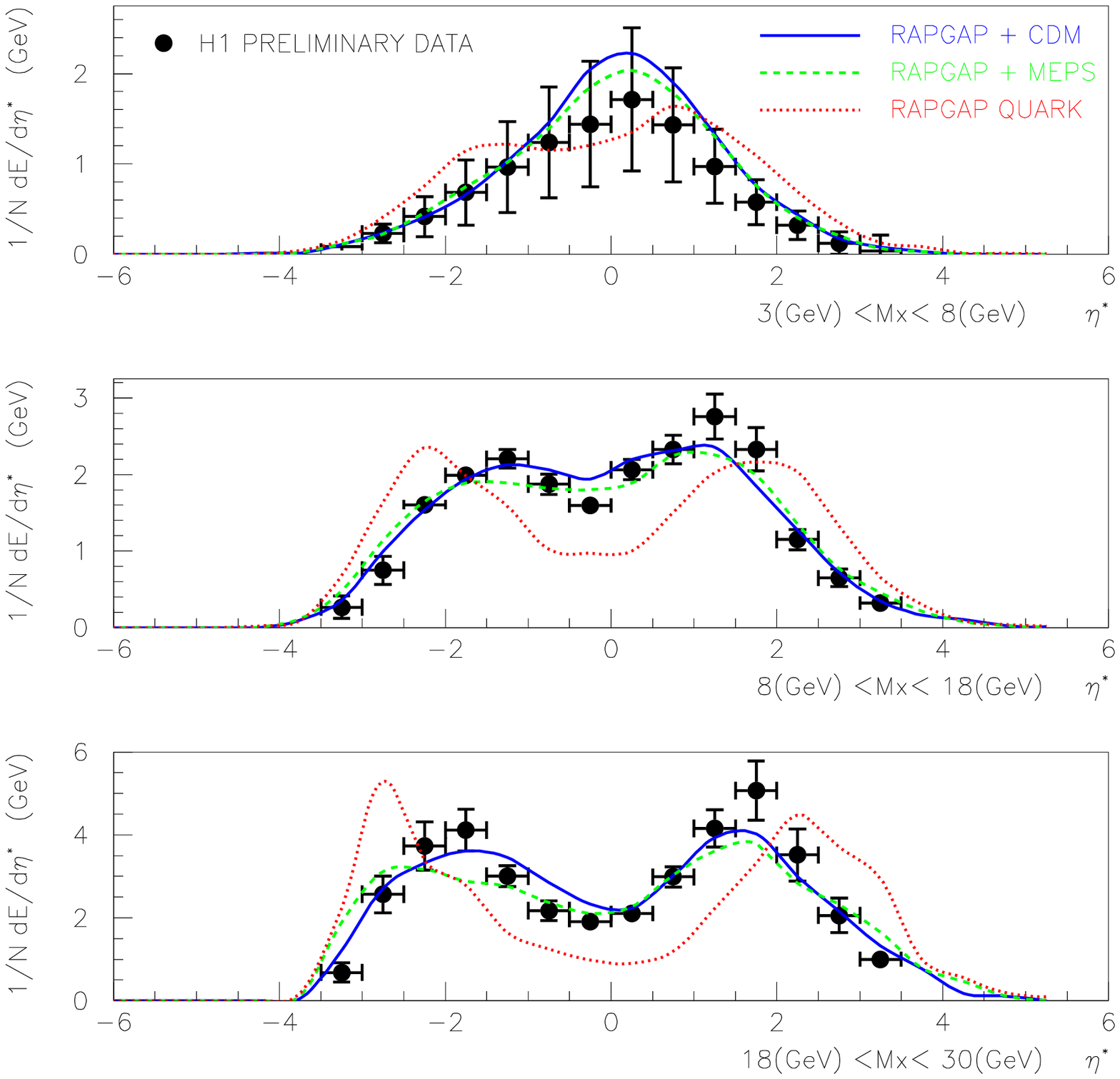,width=82mm,height=102mm}
   \put(-26,74){\epsfig{file=white2.eps}}
   \put(-24.4,88.8){\scriptsize \tt RG-QG (CDM)}
   \put(-24.4,85.8){\scriptsize \tt RG-QG (MEPS)}
   \put(-24.4,82.8){\scriptsize \tt RG-Q~~(CDM)}
  \end{picture}
  \normalsize
  \caption{Energy flows in the $\gamma^*\Pom$ centre of mass system 
           in three bins of $M_{_X}$, 
           as measured in the H1 detector~\protect\cite{H1_flow}.
           The data are compared to RAPGAP~\protect\cite{RGMC} Monte
           Carlo predictions, using either pure quark (Q), or gluon
           dominated (QG) parton densities in the Pomeron.}
 \label{fig:Eflow}
\end{figure}

These predictions are indeed supported by the data, as shown for example
in Fig.~\ref{fig:Eflow} and as discussed in more details 
elsewhere~\cite{H1_flow}.

\begin{figure*}
 \begin{picture}(170,62)(1,5)
   \epsfig{figure=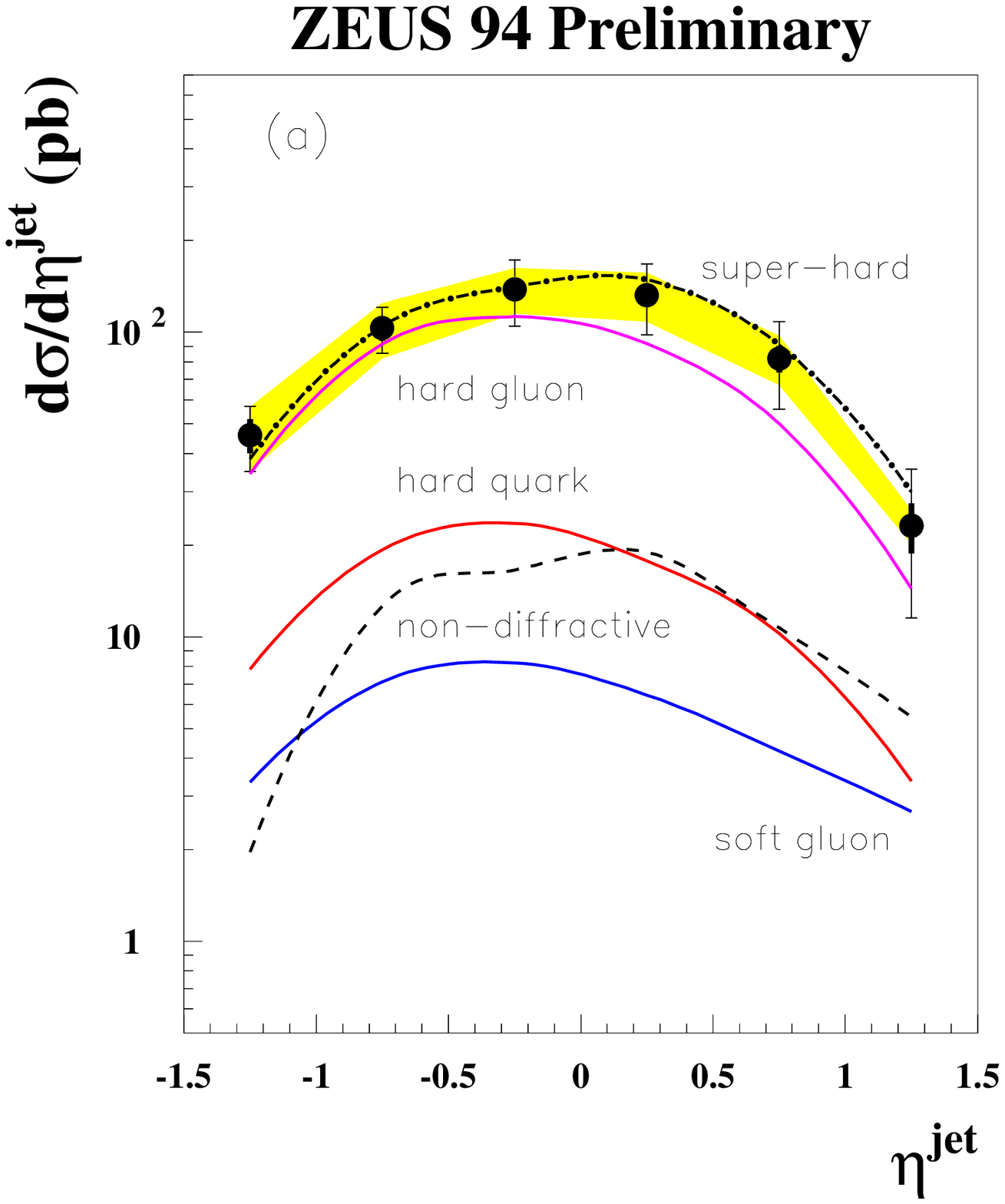,width=53mm,%
    bbllx=40pt,bblly=100pt,bburx=500pt,bbury=690pt}
   \epsfig{figure=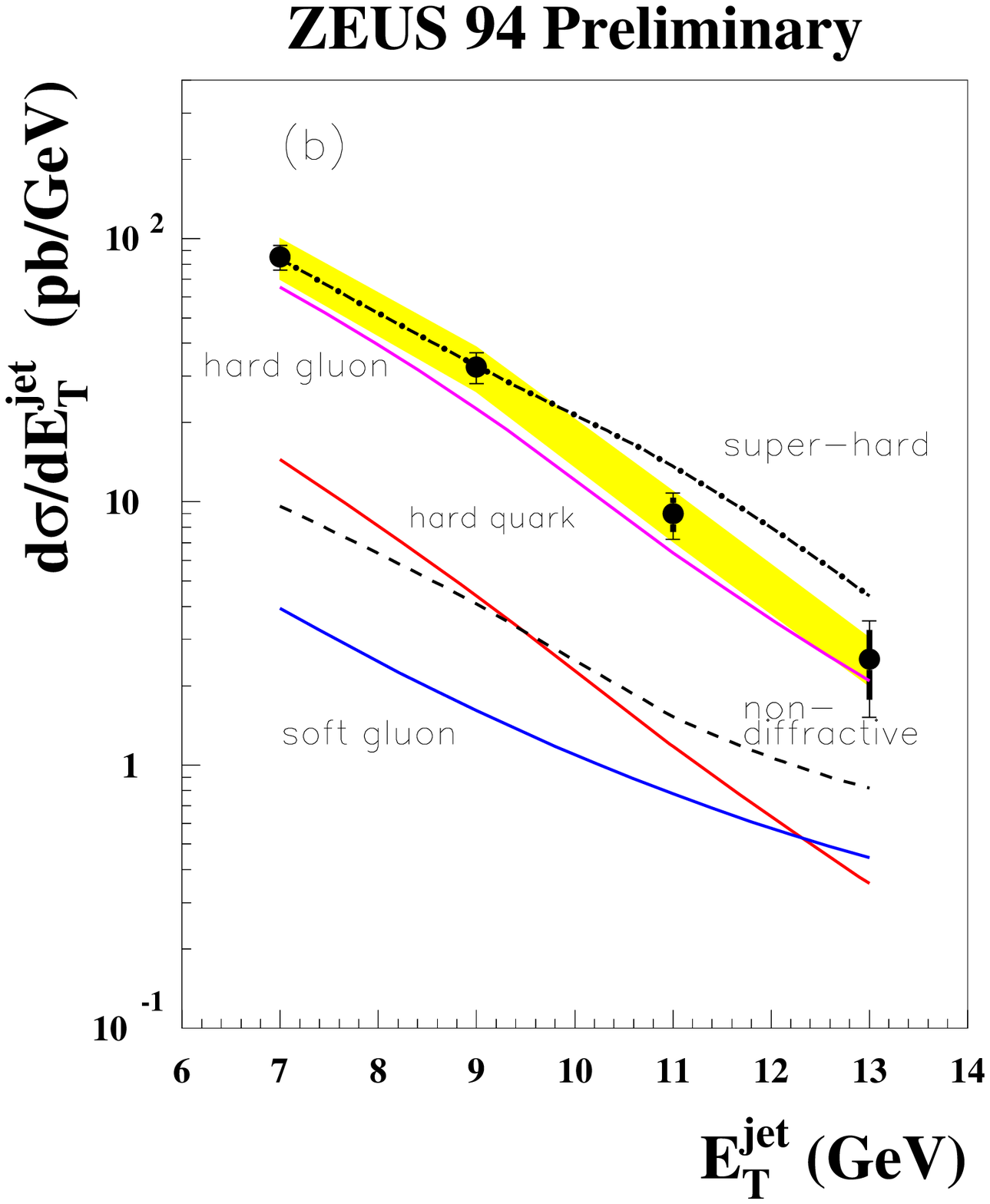,width=53mm,%
    bbllx=40pt,bblly=100pt,bburx=500pt,bbury=690pt}
   \epsfig{figure=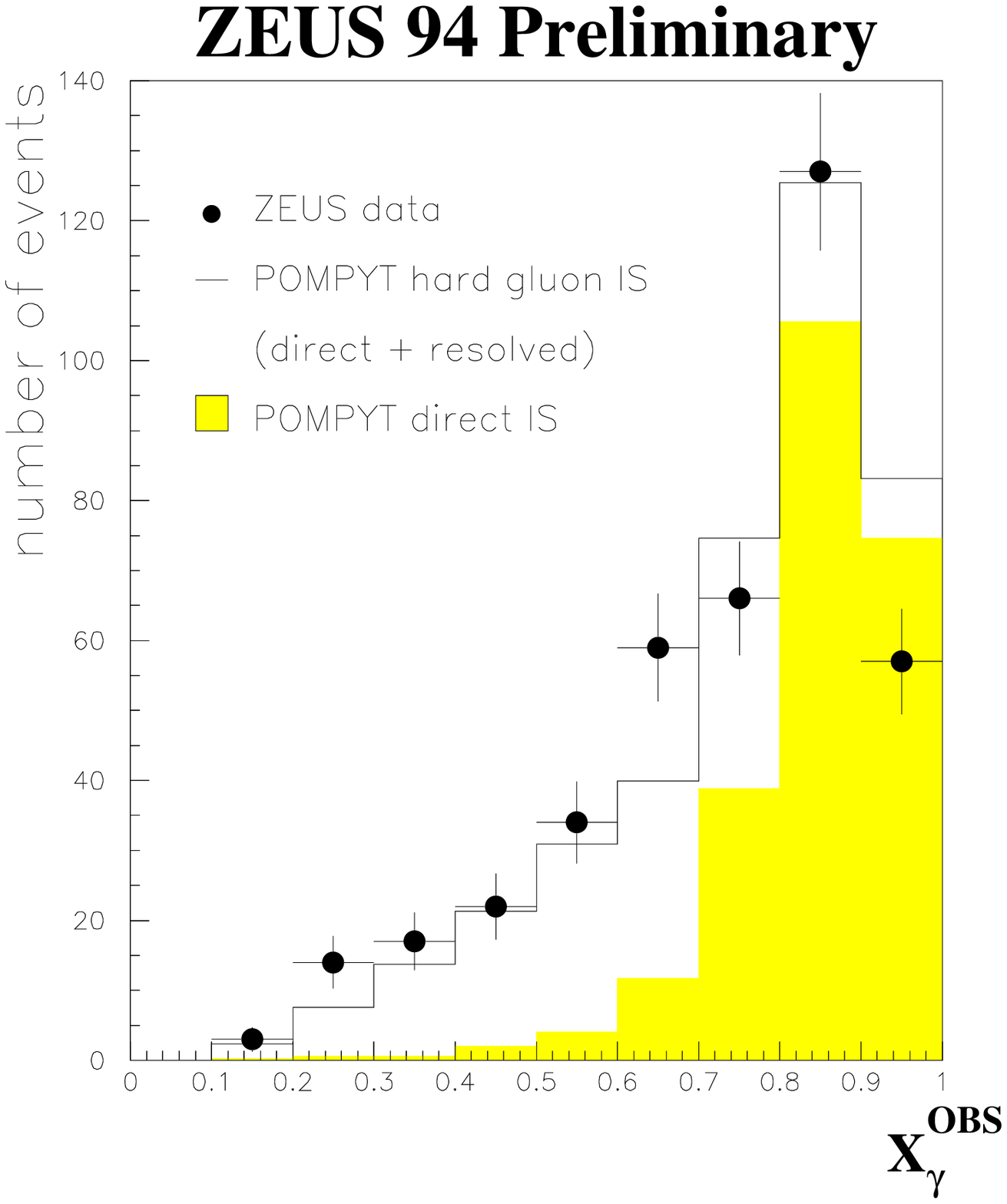,width=53mm,%
    bbllx=40pt,bblly=100pt,bburx=500pt,bbury=690pt}
   \small
   \put(-149,54){\epsfig{file=white1.eps}}  \put(-134,56){(a)} 
   \put(-093,54){\epsfig{file=white1.eps}}  \put(-079,56){(b)} 
   \put(-26,56.3){(c)}
   \normalsize
  \end{picture}
  \caption{Dijet cross sections in diffractive photoproduction at HERA,
           as measured in the ZEUS detector~\protect\cite{Z_gpj}.} 
 \label{fig:gp_jets}
\end{figure*}

The partonic structure of the Pomeron, as measured in DIS,
is also consistent, at least qualitatively,
with diffractive {\em photoproduction} of high $E_t$ jets.
This is illustrated by Fig.~\ref{fig:gp_jets}(a,b)
where ZEUS data~\cite{Z_gpj}
are seen to prefer a hard gluon density in the Pomeron.
Another interesting observation in this analysis, 
shown in Fig.~\ref{fig:gp_jets}c, is that a sizable 
proportion of diffractive hard photoproduction is produced via the so called
{\em resolved} photon contribution, in which only a fraction of the
photon momentum, $x_{\gamma}$ participates in the hard subprocess. 

The H1 measurement~\cite{H1_F2D} of the Pomeron intercept,
$$ \Delta = 1.18 \pm 0.02_{stat} \pm 0.04_{syst}, $$ 
extracted from the $\xP$-dependence of the cross section~(\ref{eq:F2D3}),
is in agreement with  the result of a similar ZEUS analysis~\cite{Z_F2D}
and is  somewhat larger, than the intercept of the `soft' Pomeron.

When interpreting these results it should be remembered, that in a fully
inclusive case, integrating over all possible diffractive final states,
both  high $p_t$ (short distance) and  low $p_t$ (long distance) 
partonic configurations contribute, 
and it is not a priori clear which ones will dominate.
There are arguments~\cite{DD_BJ} for low $p_t$ physics to prevail
even in the case of DIS diffraction.
Therefore, to isolate a possible contribution from the `hard' Pomeron,
it is necessary to select specific final states for which 
hard scale physics processes control the cross section 
(heavy quarks~\cite{Ryskin,Levin_cc}, vector mesons  produced 
by longitudinally polarized virtual photons~\cite{FR_ST}, 
high $E_t$ jets etc.)
Statistically adequate data samples for such kind of analyses,
permitting quantitative comparisons with pQCD calculations and capable
for discriminating between various models,
require higher luminosities at HERA and yet improved 
detector capabilities, which are both in progress.

\subsection{Photoproduction of Vector Mesons} \label{subsec:vm} 
Exclusive vector meson production in $ep$ collisions,
being sensitive both to the intercept $\Delta$ and to the slope $\aPP$ of
the leading Regge trajectory~(\ref{eq:sP}), provides 
yet another possibility to study the properties of the Pomeron at HERA.
A great attention to this area was attracted 
after it was realized, that in many cases pQCD is applicable allowing the
cross sections to be analytically calculated~\cite{Ryskin,FR_ST,SJB}.
It was shown, that the evaluated cross sections are proportional to the square 
of the gluon density in the target, thus providing a very sensitive measure
of the gluon distribution in the nucleon. 
In this section a brief overview is given of the experimental status of
elastic photoproduction of vector mesons at high energies.
New data on electroproduction of vector mesons at high $Q^2$ are discussed
elsewhere~\cite{GAb}.

\begin{figure}[t]
 \center
  \begin{picture}(87,30)(0,18)
  \begin{sideways} \put(31,0){\small b(GeV$^{-2}$)} \end{sideways}
  \put(54,3.8){\small $W_{\gamma p}$(GeV)}
  \put(61.5,40){(a)}
  \put(15,42.2){\small low energy data~\cite{bel_low}}
  \put(15,39.8){\scriptsize H1~\cite{bel_H1}}
  \put(15,37.4){\scriptsize ZEUS-93~\cite{bel_Z93}}
  \put(15,34.9){\scriptsize ZEUS-94~\cite{bel_Z94}}
  \put(-11,0){\epsfig{figure=F_11a.eps,width=88mm,height=57mm,angle=90}}
  \end{picture}
  \begin{picture}(80,96)(-2,1)
  \put(63,74){(b)}  \put(63,33){(c)}
  \epsfig{figure=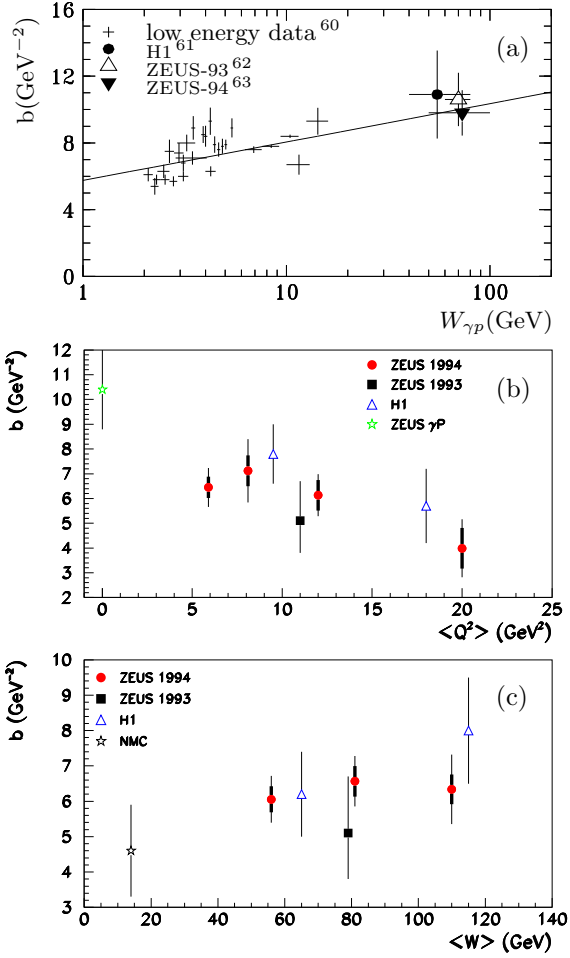,width=80mm,height=85mm}
  \end{picture}
  \caption{Energy dependence of the elastic slope in (a) photo- and
           (c) electroproduction.
           The full line  shows Regge parametrization
               $b = b_0 + 2\aPP \ln W_{\gamma p}^2$
               with $b_0 = 5.75$ and $\aPP = 0.25$;
           (b) $Q^2$ dependence of the slope of elastic electroproduction
               of $\rho^0$ mesons at HERA.}
 \label{fig:bslope}
\end{figure}

A whole bunch of fresh results~\cite{bel_Z94,Z_rho,Z_om}
on elastic and diffractive photoproduction of light mesons 
($\rho^0, \omega$) has been reported at this conference 
by the ZEUS Collaboration.
For the first time at HERA $t$-distributions were directly measured by
exploiting the Leading Proton Spectrometer of ZEUS.
The measurements of the slope $b$ of the differential cross section,
$d\sigma(\gp\!\ra\!\rho^0p)/dt = A e^{bt}$ at different energies 
are compiled in Fig.~\ref{fig:bslope}a. 
It is seen, that the shrinkage of the
`diffractive cone' is well described by the parameters of the `soft' Pomeron.
For comparison, 
the behavior of the slope $b$ as a function of $Q^2$ (Fig.~\ref{fig:bslope}b)
and $W$ (Fig.~\ref{fig:bslope}c) are also shown in case of elastic 
electroproduction of $\rho^0$.
Remembering that the slope $b$  reflects a typical interaction radius: 
$b \propto (R_p^2 + r_{\perp}^2)$, one can conclude
that the radius $r_{\perp}$ increases
with energy $W$ and decreases with virtuality $Q^2$,
in qualitative agreement with expectations.
In this somewhat simplified picture, when $r_{\perp}$ becomes much smaller
than the size of the proton, $R_p$, the slope $b$ asymptotically tends to
a constant value. This corresponds to $\aPP \approx 0$ for the `hard' Pomeron.
More precise data on $t$-distributions in elastic electroproduction of 
vector mesons are required in order to measure
the evolution of $\aPP$ with $Q^2$ at HERA.

\begin{figure}
 \center
  \begin{picture}(80,76)(0,0)
  \epsfig{figure=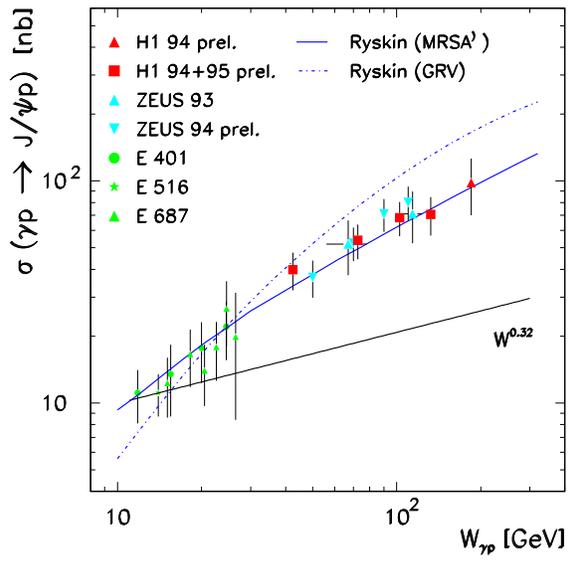,width=83mm}
  \end{picture}
  \caption{Energy dependence of elastic $J/\psi$ photoproduction.
           The line $W^{0.32}$ indicates the prediction, based on
           `soft' Pomeron.}
 \label{fig:VM1}
\end{figure}

New measurements of elastic $J/\psi$ photoproduction at HERA~\cite{Jpsi}
are compared in Fig.~\ref{fig:VM1} to low energy data and to the model 
prediction~\cite{Ryskin}, based on pQCD calculations and using two different
parametrizations for the proton structure function. 
One can see, that the rate of $J/\psi$ production grows
with energy significantly faster, as compared to the Regge inspired
expectation, $\sigma_{el} \propto W^{4\lambda}$, 
based on the `soft' Pomeron ($\lambda \approx 0.08$).
On the contrary, the measured cross section is in  good agreement with
pQCD model, showing in addition the predicted sensitivity 
to the gluon density in the proton.
One should mention, that another model~\cite{CKMT}, based upon Reggeon
field theory with $Q^2$ dependent unitarity corrections, also gives
a very good description of the data~\cite{HKK}.  
\begin{figure}
 \center
  \epsfig{figure=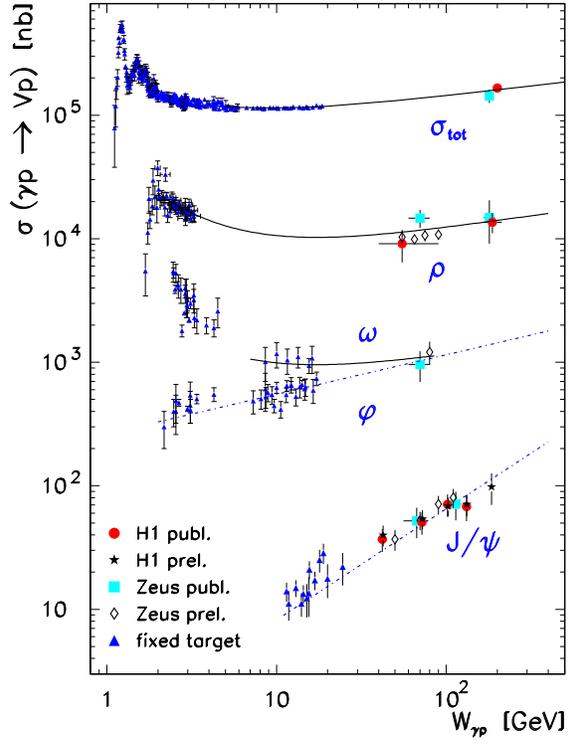,width=83mm}
  \caption{Total $\gp$ and exclusive vector meson 
           photoproduction cross sections.}
 \label{fig:VM2}
\end{figure}

The energy dependences of the total and elastic 
($\rho, \omega, \phi$ and $J/\psi$) photoproduction cross sections are 
compiled in Fig.~\ref{fig:VM2}.
A clear difference in the steepness indicates the transition from 
the `soft' to `hard' Pomeron caused by the large mass of the charm quark, 
which defines a hard scale in case of $J/\psi$ production.

\subsection{Concluding Remarks} \label{subsec:concl} 
To summarize,
both soft and hard diffractive processes have been observed at HERA,
in real and virtual photon-proton collisions.
The data reveal characteristic features of 
either a {\em soft} (non-perturbative), or a {\em hard} 
(perturbative) Pomeron, depending on the value of the typical scale
involved, in qualitative agreement with theoretical expectations.
The properties of diffractively produced  final states  in
photoproduction as well as in the DIS regime are consistent with `leading'
gluon component in the Pomeron.
\begin{figure}[h]
 \center
  \begin{picture}(70,90)(21,0)
   \begin{sideways} 
    \put(67,-11){DL~\cite{d_DL}} \put(63,-21){Tevatron~\cite{d_TevP}} 
   \end{sideways}
   \put(47,60.0){$\sigma_{tot}(\gamma p)$}
   \put(47,51.3){$\gamma p \ra \rho^0 p $}
   \put(47,43.0){$\gamma p \ra M_x p$}
   \put(47,34.3){$\gamma^*p \ra M_x p$} 
   \put(47,26.5){$\gamma p \ra J/\psi p$}
   \put(47,17.2){$\gamma^*p \ra \rho^0 p$}
   \put(56,79){H1}  \put(56,75.1){ZEUS} 
   \put(56,71.2){HERA}  \put(56,67.7){world}
   \put(31,0){\large $\alpha_{\Pom}(0)$}
   \epsfig{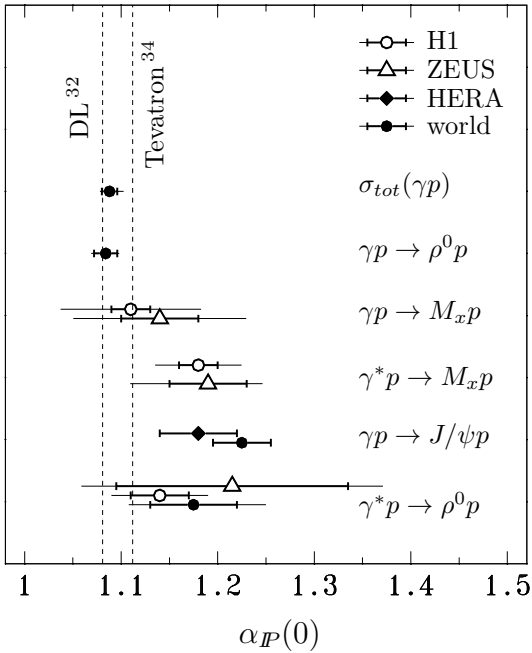}
  \end{picture}
  \caption{Summary on the Pomeron intercept measurement at HERA.
           Thick inner bars represent statistical errors, while the total
           (thin) bars correspond to statistical and systematic errors
           added in quadrature.
           Dashed lines show the values obtained from the combined fit 
           of the total hadronic cross sections~\protect\cite{d_DL}
           and from the Tevatron data~\protect\cite{d_TevP}.}
 \label{fig:PI}
\end{figure}

A summary of the Pomeron intercept measurements at HERA is presented
in Fig.~\ref{fig:PI}. 
Although no evidence is found for the naive leading order BFKL 
Pomeron~(\ref{eq:BFKL}), the data clearly show the departure from the
`soft' Pomeron towards larger values of $\aP(0)$, whenever 
a {\em hard scale} is present.
More precise data coming from new HERA run are expected to further 
clarify  this intriguing situation.

\section{From Soft to Hard Physics} \label{sec:four} 
One of the `goldplated' fields of research at HERA is
to study how the properties of photon and the dynamics of
its interaction with protons
evolve with $Q^2$, by comparing real and virtual photon-proton 
interactions in identical experimental conditions.
Interesting information can be extracted from both 
inclusive measurements and more detailed properties of the hadronic
final states.
A review of recent ideas in modelling the transition from
photoproduction to DIS regime can be found elsewhere~\cite{Trans}.

New measurements of the proton structure function $F_2$ at HERA,
presented at this conference~\cite{F2_lowq}, have demonstrated  that
perturbative QCD successfully describes the data down to surprisingly
low $Q^2 \simeq 2$ GeV$^2$. The transition between `soft' ($\lambda
\approx 0.08$) and `hard' ($\lambda \geq 0.2$)  regimes occurs, at HERA
energies, at $Q^2 \approx 1$ GeV$^2$.

\begin{figure}[h]
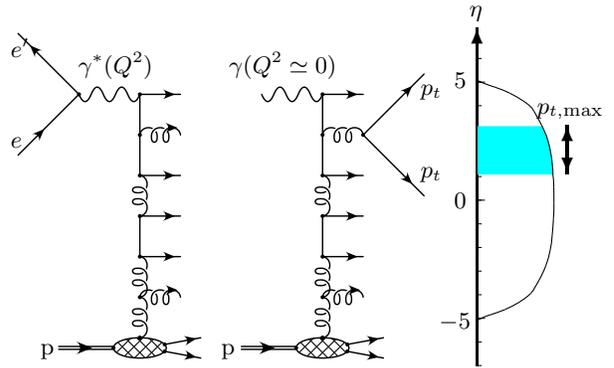

 \center
 \begin{picture}(80,50)(-2,-3)
  \thicklines   \small
  \put(08,38){$\gamma^* (Q^2)$}  \put(28,38){$\gamma (Q^2 \simeq 0)$}
  \put(-1,28){$e$}               \put(-1,40){$e'$}  
  \put(3,0.5){p}                 \put(27,0.5){p}
  \put(53.5,24){$p_t$}           \put(53.5,35){$p_t$}
  \put(58,36){5}  \put(58,20){0} \put(56,04){$-$5} 
  \put(60,45.5){$\eta$}          \put(69,32.5){$\ptm$}
  \put(61,-1){\vector(0,1){45}}  \put(73,27){\vector(0,1){4}}  
  \put(73,27){\vector(0,-1){2.5}}
  \normalsize
  \put(61,-1){\epsfig{file=F_15b.eps,height=44mm}}
  \epsfig{file=F_15a.ps,width=54mm}
  \end{picture}
  \caption{Schematic diagrams for DIS event and photoproduction event
           with high $p_t$ particles in the final state. 
           The sketch on the right side illustrates 
           an inclusive charged particle pseudorapidity distribution
           in $\gamma p$ centre of mass system, with the range 
           used to search for high $p_t$ tracks in the H1 detector.}
 \label{fig:Tr1}
\end{figure}

Another way to study similarities and differences between photoproduction
and DIS has been presented by the H1 Collaboration~\cite{H1_Wdep}.
The idea of the analysis, as illustrated in Fig.~\ref{fig:Tr1}, is to use
high $p_t$ charged particles as a measure of a typical scale in $\gp$
process, which can be varied in a similar way as a scale $Q^2$ in DIS.

\begin{figure}
 \center
  \begin{picture}(85,57)(4,0)
   \small                                                      
   \begin{sideways} \put(44,-3){slope $\lambda$} \end{sideways}  
   \put(17.5,50.2){H1 $\gamma p$ data}
   \put(17.5,46.7){Phase space}       
   \put(17.5,43.2){PYTHIA} 
   \put(51.5,50.2){H1 DIS data}
   \put(51.5,46.7){GRV94} 
   \put(51.5,43.2){MRS-D0$^{\prime}$}
   \put(37,47){(a)} \put(70,47){(b)} 
   \normalsize
   \epsfig{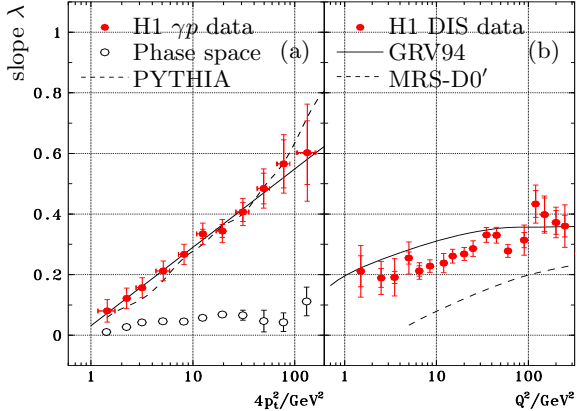}
   \end{picture}
   \caption{The scale dependence of the slope $\lambda$ in 
            photoproduction (a) and DIS (b). 
            The full line in (a) represents the result of a linear fit 
            $\lambda_{\gp}=C\cdot\ln(4p_t^2)+\lambda_0$ to the data, 
            with $C=0.112\pm0.007$ and $\lambda_0=0.031\pm0.014$.
            The PYTHIA prediction is based on Monte Carlo 
            version~\protect\cite{PYTHIA}
            which includes both pQCD for the hard scattering process 
            and a Regge inspired soft interaction component.} 
  \label{fig:Tr2}
\end{figure}

Good detector acceptance and high resolution of $p_t$ measurement, 
together with large statistics of $\gp$ events, allow a continuous 
coverage of both soft and hard scattering domains.
Similar to $\sigma_{\gamma^*p}(W,Q^2) \propto W^{2\lambda(Q^2)}$ in DIS,
the photoproduction cross section was parametrized as
$\sigma_{\gp}(W,\ptm^2) \propto W^{2\lambda(\ptm^2)}$ with the 
scale defined by the maximum $p_t$ charged particle in the photon
fragmentation region.
The results of this analysis are shown in Fig.~\ref{fig:Tr2}.
One can see a clear similarity in the evolutions of the $W$ dependence
of $\gp$ and $\gamma^*p$ cross sections with increasing scales
($4\ptm^2$ and $Q^2$ respectively).
Another interesting observation is that the transition from soft to hard
regimes proceeds in both cases smoothly. 
   
This similarity can be qualitatively understood both in terms of pQCD and
in terms of a Regge approach.
In perturbative QCD the effect of more steeply rising cross sections
with increasing hard scale can be readily explained 
by leading order partonic scattering contributions~\cite{Halzen,Haidt},
while in Reggeon Field Theory (RFT) it is a consequence of the reduction
of unitarity corrections to one-pomeron exchange~\cite{CKMT,Levin}.

It is informative to compare also the properties of the hadronic final
states in photoproduction and DIS.
As was explained in Section~\ref{subsec:LRG} experimental conditions 
at HERA allow the measurements in the central rapidity plateau 
in the $\gamma^*p$ centre of mass system,
and are especially favorable to study the photon fragmentation region.

From simple, but very general arguments~\cite{tr_BJ} the central region
is expected to depend only on the total energy, 
but not on the type of colliding particles. 
This was verified in hadron-hadron collisions.
All the difference between real and virtual photon-proton interactions
is expected to be seen in the photon fragmentation region:
at high $Q^2$ the  situation should resemble the \eea~ case,
whereas  final states produced by real photons are expected to be similar
to those in hadron-hadron collisions.
In particular, one expects transverse energy flow, particle
multiplicity and $\langle p_t \rangle$ to grow with $Q^2$.

First analysis of HERA data along this line~\cite{H1_eflow}, 
in which energy flow was compared in $\gp$ and DIS reactions, 
confirmed the picture.
At this conference new preliminary analysis has been reported~\cite{H1_nch}
in which charged particle spectra in pseudorapidity, $dN_{ch}/d\eta$, 
are studied in two data samples: 
$Q^2 \approx 0$ and $\langle Q^2 \rangle = 15$ GeV$^2$, at 
$\langle W \rangle = 190$ GeV.

\begin{figure}[hb]
 \center
  \begin{picture}(100,53)(22,9)
   \put(46,57){\bf H1 PRELIMINARY}
   \put(52,48){a)} \put(93,48){b)} \put(55,7){$\eta$} \put(96,7){$\eta$}
   \put(35,24.5){$p\bar{p}$ (UA5)} \put(35,17.9){$\gamma p$ (H1)}
   \put(74.5,24.0){$\gamma^*p~(\overline{Q^2}=15)$}   
   \put(74.5,17.8){$\gamma p~~(Q^2=0)$}
   \begin{sideways} 
    \put(40,-11){$\overline{dN_{ch}}/d\eta$} 
   \end{sideways}
   \epsfig{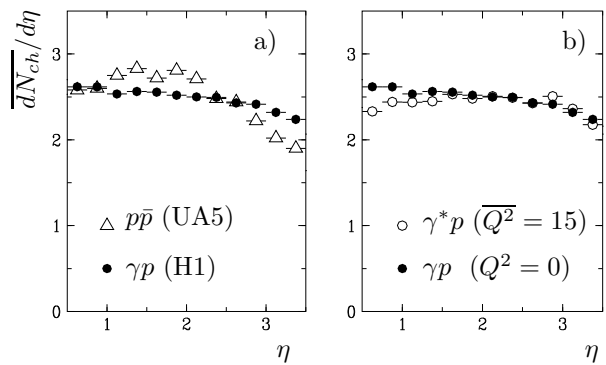}
  \end{picture}
  \caption{Charged particle pseudorapidity distributions in 
           non-diffractive interactions of \pp at $ E_{cm} = 200$ GeV
           and $\gamma p$, $\gamma^*p$ at $W=190$ GeV.}
 \label{fig:Tr3}
\end{figure}

In Fig.~\ref{fig:Tr3} the data from non-diffractive $\bar{p}p$~\cite{UA5},
$\gp$ and $\gamma^*p$ interactions are compared.
Surprisingly, we discover, that the difference between similar `soft'
reactions (\pp and $\gp$) is even bigger than between $\gp$
and DIS samples.  
This means that the whole difference in charged particle multiplicity
between photoproduction and DIS
in the photon fragmentation region is related entirely to the
different proportion of diffractive events in the two reactions.
This somewhat unexpected result remains to be understood and explained.

Apart from the already discussed motivation, it is interesting to test
a remarkable property of RFT, that the unitarity correction contributions
cancel in inclusive particle cross sections
in the central rapidity region.
This is a direct consequence of AGK cutting rules~\cite{AGK}, which relate
multi-pomeron exchange diagrams of RFT to particle production. 
Hence, within this framework the behavior of
\beq
    \left. \frac{d\sigma_{ch}}{d\eta}\right|_{\eta \approx 0}
              \stackrel{\mbox{\scriptsize AGK}}{~=~} 
    \left. \sigma^{(\!1\!)} \frac{dN_{ch}^{(\!1\!)}}{d\eta}
                  \right|_{\eta \approx 0}
\label{eq:AGK} 
\eeq
is predicted to be universal in all types of high energy reactions
and is defined by the properties of the `bare' one-pomeron exchange,
denoted in eq.~(\ref{eq:AGK}) by superscript~$^{(1)}$.

\begin{figure}
 \center
  \begin{picture}(95,54)(26,1)
   \put(88.3,1){$W$ [GeV]} 
   \put(48,48.3){$\sim W^{2\cdot(0.031\pm0.006)}$}
   \put(48,43.3){$\sim W^{2\cdot(0.105\pm0.004)}$}
   \put(40.5,18.8){DIS (EMC~\cite{EMC})} \put(40.5,14.0){DIS (H1 prelim.)}
   \put(74.5,19.2){$\gamma p$ (H1 prelim.)}   
   \put(74.7,14.5){$pp$~\cite{ISR,NA22}, $p\bar{p}$~\cite{UA5}}
   \begin{sideways}  \put(45,-13.5){$\rho(0)$} \end{sideways}
   \epsfig{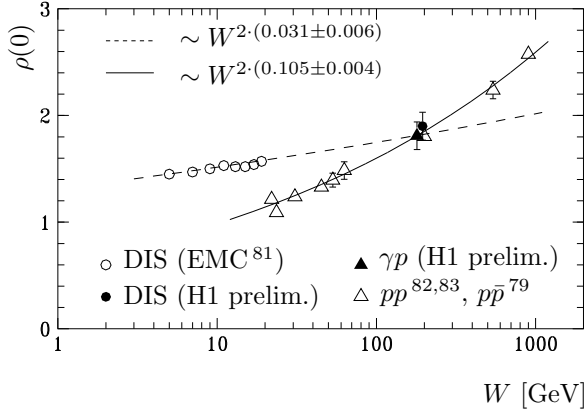}
  \end{picture}
  \caption{Energy dependence of the charged particle density in the central
           rapidity region in different reactions.}
 \label{fig:Tr4}
\end{figure}

Fig.~\ref{fig:Tr4} shows the energy dependence of the charged particle
density
$$  \rho(0) = \frac{1}{\sigma_{tot}} \frac{d\sigma_{ch}}{d\eta} $$ 
in the central rapidity region in DIS, photoproduction and 
$hp$ interactions.
It is observed that the energy dependence of DIS data is significantly
weaker.
This is in agreement with prediction~(\ref{eq:AGK}), and is explained
by the difference in the energy dependence of $\sigma_{tot}(\gp/hp)$ 
and $\sigma_{tot}(\gamma^*p)$.

To further test the validity of RFT predictions, it would be interesting
to compare the shapes of MD and two-particle correlation functions
for which differences are expected between photoproduction and DIS.

\section{Summary}
Soft processes represent an essential part of high energy physics.
An ultimate goal here is to gain analytic control over confinement
and thus to complete QCD as a theory of strong interactions.
The task is difficult, but worth the effort.
All experimental facilities are important, providing complementary
information in this field of research.

Recent experimental data have demonstrated that perturbative QCD works
extremely well down to surprisingly low scales, both in the description
of time-like partonic cascades in \eea~ and 
space-like cascades in deep inelastic scattering.\footnote
   {In that respect, an interesting analogy has been noted~\cite{ADR_EDW}
    at this conference,
    between the energy dependence of the charged particle multiplicity
    in \eea~ and the rise of $F_2$ at low $x$ at HERA.
    The theorists are invited to judge whether this is just a lucky chance,
    or a  natural feature of pQCD dynamics.
    In any case, an amazing consequence is that one could predict
    low $x$ behavior of the $F_2$ by measuring 
    $\langle n_{ch}\rangle_{e^+e^-}(s)$ and vise versa!}
 
Interesting qualitative similarities have been observed between 
hard processes (\ee an\-ni\-hi\-lation, DIS) 
and those which are predominantly driven by soft physics 
($hh$ collisions, $\gp$ scattering).
This can provide a good basis for understanding the relations between
Regge theory and QCD, as well as underlying dynamics in the transition region 
between the two.

One of the best opportunities to investigate an interplay between soft
and hard physics is offered by diffractive deep inelastic scattering.
For the first time we have a powerful facility for detailed study of the
partonic content of diffractive exchange.
Yet more precise data from HERA, together with improving detector 
capabilities promise that an interesting life in this field will continue,
challenging the theorists and providing new insights into the properties
of the Pomeron both in perturbative and non-perturbative regimes.

\section*{Acknowledgements}
It is a pleasure to thank Professor A.K. Wroblewski and the Organizing
Committee for the kind invitation to give a review talk at this conference.
I am also grateful to my colleagues from H1, ZEUS, OPAL, DELPHI
and L3 for providing me with their data and for many useful discussions.
Special thanks go to John Dainton for taking a critical look at the 
manuscript.

\section*{References}


\end{document}